\documentclass[RNAAS]{aastex631}
\usepackage{mathtools}
\usepackage{amsmath}
\usepackage{CJKutf8}
\usepackage{xcolor}
\usepackage{natbib}

\newcommand{\reff}{$R_{\rm eff}$}

\begin{document}

\begin{CJK*}{UTF8}{gbsn}
\title{A Visually Classified Spectroscopic Object Catalog for SDSS-IV MaNGA}

\author{Joshua~L.~Steffen}
\affiliation{Department of Physics \& Astronomy, University of Iowa, Iowa City, IA 52242}
\author{Hai~Fu}
\affiliation{Department of Physics \& Astronomy, University of Iowa, Iowa City, IA 52242}

\begin{abstract}

We provide a catalog of visually classified objects in the MaNGA integral field spectroscopic survey. The MaNGA survey is designed to target a single galaxy with each of its integral field units (IFUs); however, many of these fields will host ancillary objects. We identify these discrete objects by cleaning up SDSS photometric objects in MaNGA's fields-of-view. We then use the spectra from MaNGA's data-cubes to spectrally classify the identified objects. The catalog contains the positions and classifications of 1385 stars, 11,439 galaxies, and 107 broad-line AGN (BLAGN) from the 10,130 unique MaNGA fields. We also provide spectroscopically derived parameters for the galaxies including; stellar masses, gas and stellar kinematics, and emission-line fluxes and equivalent widths. This catalog effectively expands the size of the MaNGA catalog by $\sim$50\%, increasing the utility of the MaNGA project.

\end{abstract}


\section{Introduction}\label{sec:intro}
MaNGA (Mapping Nearby Galaxies at Apache Point Observatory) is a massive integral field spectroscopic survey of nearby galaxies. The project feeds 17 IFUs (optical fiber bundles; \citet{Law15}) into two dual-channel BOSS spectrographs \citep{Drory15} on SDSS's 2.5 meter telescope. The survey targets nearby galaxies within a redshift range of 0.01 $<$ $z$ $<$ 0.15 and a luminosity range of -17.7 $<$ $\mathcal{M}_{\rm i}$ $<$ -24.0, where $\mathcal{M}_{\rm i}$ is the rest frame i-band absolute magnitudes from elliptical Petrosian apertures. MaNGA's final data release (DR17; \citet{Abdurrouf21}) contains 10,130 unique observations. 

Each IFU observation is designed to cover a single galaxy out to 1.5 \reff\ and 2.5 \reff\ (where \reff\ is the half light radius). The extended spatial coverage was intended to be used to study galaxy properties across the spatial dimensions, but the it also allows many ancillary objects to fall into MaNGA's fields-of-view. Here, we provide the first catalog that identifies and classifies all spectroscopic objects within MaNGA's 2392 arcmin$^2$ footprint.

\section{Methods}\label{sec:method}

\begin{figure}
\centering
\includegraphics[width=5in]{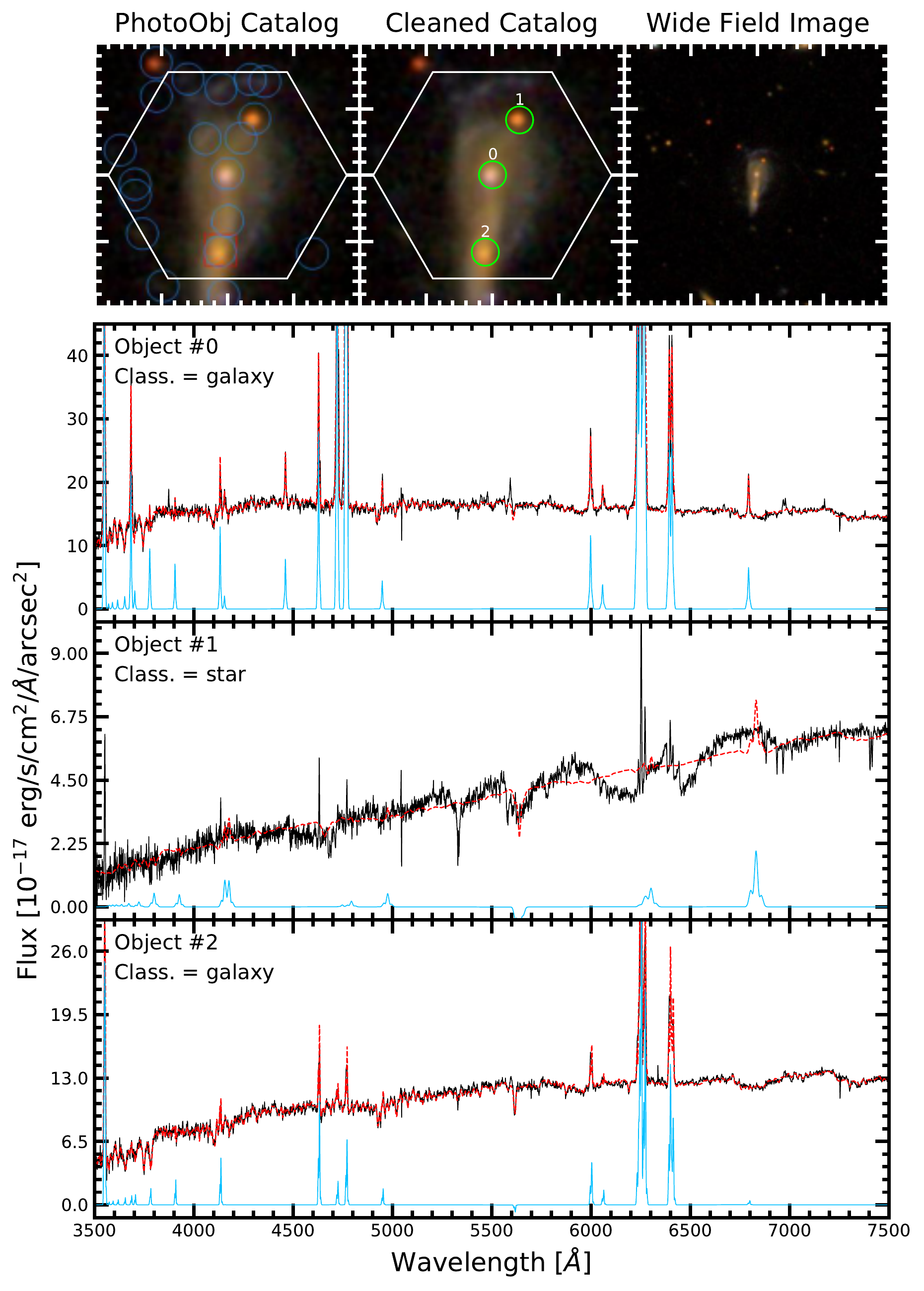}
\caption{A example of our identification and classification pipeline for a single MaNGA field (Plate-IFU: 12512-12702). The top row shows SDSS pseudocolor images for the MaNGA field. The image on the left shows the 40$^{\prime\prime}$ field with SDSS's {\sc photoObj} catalog overlaid in blue circles and SDSS's spectroscopic catalog overlaid in red squares. The white hexagon represents MaNGA's field-of-view. The image in the middle shows the same field but with the cleaned catalog in green circles. The numbers above the green circles are object IDs. The right image shows the same object with a wider field (160$^{\prime\prime}$ across). The bottom three rows show the spectra for each object. The black spectra represents the extracted spectra from MaNGA's data-cubes, the red spectra is SPFIT's best-fit model spectra, and the blue spectra are the best-fit emission lines. In the top left corner of each spectra plot shows the object ID and our spectral classification. }
\label{fig:fig}
\end{figure}

The construction of our MaNGA object catalog consists of two main stages. We first identify objects in MaNGA's fields-of-view by cross-matching the survey with previous catalogs. Next, we use the spectra provided in MaNGA's data-cubes to spectroscopically classify the objects. Along with the spectral classification, we can use the best-fit spectral models to extract observed and derived properties of these objects.

Properly deblending discrete objects in crowded fields is a challenge. Some algorithms under-deblend imaging data such that it misses sources in crowded fields. Other algorithms over-deblend fields such that several objects are assigned to a single extended object. MaNGA's targeting catalog, the NASA Sloan Atlas\footnote{NSA v1\_0\_1; \url{http://www.nsatlas.org}} (NSA), frequently combines objects in crowded fields while SDSS's photometric catalog, {\sc photoObj}\footnote{\url{https://www.sdss.org/dr17/imaging/catalogs/}}, frequently places extra sources over galaxies with clumpy light distributions. Instead of attempting to create our own deblending algorithm, we decide to remove false sources from the {\sc photoObj} catalog in the MaNGA fields.

We visualize the MaNGA fields by overlaying the {\sc photoObj} catalogs over SDSS pseudocolor images. We then remove these false sources by visual inspection. There are also some objects that are absent in the {\sc photoObj} catalog; we add the positions of these objects into our catalog. We also corrected a few cases in which the given positions are off when compared to the light distribution. We show an example of the object identification process in the top row of Figure \ref{fig:fig}.

SDSS's {\sc photoObj} table separates stars from galaxies based on photometry. With MaNGA, we can perform a more rigorous classification scheme using the spectra from the survey's data-cubes. We extract the spectra for every identified object with a 1$^{\prime\prime}$ radius circular aperture. We then fit the spectra with the spectral fitting algorithm, {\sc spfit}\footnote{\url{https://github.com/fuhaiastro/spfit}} \citep{Fu18}. 

{\sc spfit} is designed to simultaneously fit the stellar continuum and emission lines following the Penalized Pixel-Fitting method (\citet{Cappellari04, Cappellari17}) using the simple stellar population models from MIUSCAT \citep{Vazdekis12}. We use the redshift of the MaNGA target, from the NSA catalog, as the initial guess for the redshift for all objects in a given MaNGA field. {\sc spfit} allows doppler shifts to the model templates, but it will fail to model objects with substantially different redshifts ($\Delta v \ge$ 2000 km s$^{-1}$). 

We visually classify the identified objects by comparing the observed spectrum with the best-fit model. Since {\sc spfit} has a redshift tolerance of $\Delta v \ge$ 2000 km s$^{-1}$, we first sort objects in two redshift based categories; z\_corr and z\_off. Objects with the z\_corr designation have redshifts that are comparable to the MaNGA target while objects with the z\_off designation have a redshift that is notably different from the MaNGA target. Within the z\_corr designation, objects are classified as galaxies or BLAGN. Within the z\_off designation, objects are classified as galaxies, BLAGN, or stars. Stars, galaxies, and BLAGN can be distinguished based on their distinct spectral features. Finally, objects whose signal-to-noise is to low to discern any spectral features are classified as ``low signal" and objects with missing spectra (from being near the edge of an IFU) or defective spectra are classified as ``defective." {\sc spfit} is designed to model galaxy spectra near the input redshift, so projected galaxies, BLAGN, and foreground stars will not be properly modeled by the fitting algorithm. We show an example of the extracted and modeled spectra in Figure \ref{fig:fig}. Object \#0 and Object \#2 have galaxy type spectra while object \#1 has the spectra of a foreground star. 

\section{Results}

By visually cleaning the SDSS {\sc photoObj} catalog, we identify 15,592 objects (including MaNGA targets) within the 10,130 unique MaNGA fields. We then use MaNGA's spectra to visually classify each object and find 11,439 galaxies (of which 10,988 are z\_corr and 451 are z\_off), 1385 stars, 107 BLAGN (of which 84 are z\_corr and 23 are z\_off), 2367 low signal observations, and 294 defective spectra. 

Our spectral fitting code, {\sc spfit}, computes a number of derived properties from the modeled galaxy spectra. {\sc spfit} provides emission line fluxes, equivalent widths, and gas kinematics calculated from the fitted emission lines and stellar kinematics and stellar masses from the stellar continuum. Because the properties are derived from spectra extracted from a 1$^{\prime\prime}$ radius circular aperture centered on each object, they are only for the centers of the galaxies. We provide the positions, classifications, and derived parameters in the tables on our Github (\url{https://github.com/jlsteffen/MaNGAObj}). Since {\sc spfit} is designed to model galaxy spectra, the non-galaxy objects will have null values for the derived parameters. 

\section{Applications}\label{sec:app}

Previous versions of this catalog have already been used for studying kinematic galaxy pairs in MaNGA \citep[][Steffen et al. 2022, in preparation]{Fu18, Steffen21}, but this catalog has many other potential applications. For instance, it may be used to study stars or BLAGN with MaNGA's spectroscopy, to study galaxy clusters, or to mask ancillary objects out of the MaNGA fields. 

\begin{acknowledgments}

We acknowledge support from the National Science Foundation grants, AST-1614326 and AST-2103251. 

\end{acknowledgments}
\vspace{5mm}
\facility{Sloan}

\end{CJK*} 
\bibliographystyle{aasjournal}

\bibliography{bib}

\end{document}